\documentclass[final,1p,times]{elsarticle} 
\usepackage{graphicx} 
\usepackage{amssymb} 
\usepackage{amsthm} 
\usepackage{lineno} 

\newcommand{\Jpsi}{$J/\psi$}
\newcommand{\pT}{$p_T$}
\newcommand{\sNN}{$\sqrt{s_{\mathrm{NN}}}$ }

\newcommand{\raa}{$R_{AA}$}

\journal{Nuclear Physics A} 
\begin{document} 

\begin{frontmatter} 


\title{\Jpsi\ production in Au+Au and Cu+Cu collisions at \sNN = 200 GeV at STAR}

\author{Daniel Kiko\l{}a$^{a, b}$ for the STAR Collaboration}

\address[a]{Lawrence Berkeley National Laboratory, Berkeley, CA 94720, USA}
\address[b]{Faculty of Physics, Warsaw University of Technology, Warsaw, Poland}

\begin{abstract} 

\Jpsi\ production is considered to be a sensitive probe of the properties of quark gluon plasma created in nucleus+nucleus collisions at RHIC. In this article, the recent analysis of mid-rapidity ($\left|y \right| < 1$) \Jpsi\ production via the dielectron decay channel in Au+Au (year 2007) and Cu+Cu (year 2005) collisions at \sNN = 200 GeV at STAR is reported. It is compared to STAR p+p results in order to study the nuclear modification factor as a function of transverse momentum and centrality. The results are compared to previously published data and available theoretical models.
\end{abstract} 

\end{frontmatter} 



\section{Introduction}
Suppression of \Jpsi\ production due to screening of the binding potential of charm and anti-charm quarks in hot and dense matter is a classic signature of the quark gluon plasma formation \cite{MatsuiSatz86}  and has been intensively studied in previous years. The NA50\cite{NA501997,NA502004} and NA60\cite{Na602007} experiments at CERN-SPS (at \sNN = 17.3 GeV) observed a strong suppression of \Jpsi\ as a function of centrality. The results from the PHENIX collaboration at RHIC show that the \Jpsi\ suppression for $p_T< $ 5 GeV/c at mid-rapidity is similar to that observed at SPS energies\cite{PhenixJpsiAuAu}, although the energy density and temperature reached at RHIC are much higher than at SPS. On the other hand, the suppression at forward rapidity is stronger than at mid-rapidity. Such a pattern suggests that additional processes other than \Jpsi\ dissociation take place, such as recombination of charm quarks, sequential dissociation of exited charmonia states or feeddown from B-mesons, may compensate for \Jpsi\ dissociation. 

One of the interesting developments is the so called hot wind dissociation\cite{RajagopalWiedmann07}. Using AdS/CFT duality, it was predicted that the \Jpsi\ effective dissociation temperature decreases with increasing \Jpsi\ velocity. At RHIC energies, the expectation from the hot wind model is that the system temperature should be below the \Jpsi\ effective dissociation temperature for \Jpsi\ with $p_T <$ 5 GeV/c, and above for higher \pT, and therefore the suppression is expected to be stronger at high \pT. These predictions can be tested with \Jpsi\ measurements at high transverse momentum. 

In addition to the study of \Jpsi\ interactions with the medium produced in nuclear collisions, charmonium production in proton collisions is of interest in its own right. The production is not well understood as several mechanisms can play a role. These include direct production via gluon fusion, whether the state is formed through color singlet or octet, the contributions from parton fragmentation, and feeddown from other charmonium states and B-mesons. At the moment no model explains the major features of the existing data\cite{Lansberg:JpsiProd2006,Lansberg:JpsiProd2008}. The measurement of high-\pT\ \Jpsi\ in nucleus-nucleus collisions, together with high-\pT\ \Jpsi\ p+p data, might shed some light on charmonium production. 

In this paper, the recent studies of \Jpsi\ production in Au+Au and Cu+Cu from the STAR experiment are presented. We report and discuss the measurements of the nuclear modification factor as a function of transverse momentum and centrality. 

\begin{figure}
\begin{center}
\includegraphics[trim = 10mm 110mm 10mm 10mm, clip, scale=0.44 ]{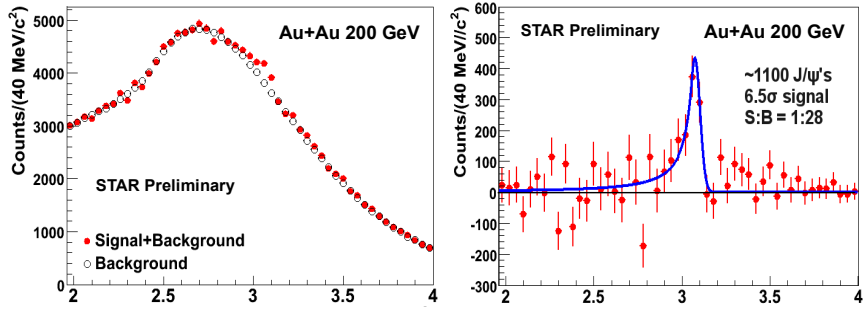} 
\caption{Left: dielectron invariant mass distribution in Au+Au \sNN = 200 GeV before background subtraction (background was estimated using event mixing techniques). Right: the $e^{+}e^{-}$ invariant mass distributions in Au+Au \sNN = 200 GeV after background subtraction; the shape of the \Jpsi\ mass peak is well described by \textit{Crystal Ball} parametrization\cite{CrystalBallF}.}
\label{JpsiSignal}
\end{center}
\end{figure}

%

\begin{figure}
\begin{center}
\includegraphics[ trim = 0mm 0mm 0mm 12mm, clip, scale=0.33]{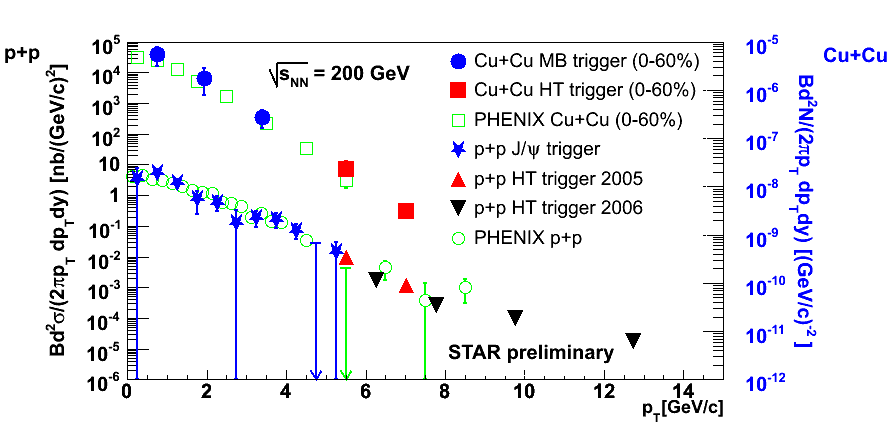} 
\caption{ \Jpsi\ $p_T$ spectrum at Cu+Cu and p+p collisions at \sNN= 200 GeV. }
\label{pTspectrum}
\end{center}
\end{figure}

\begin{figure}
\begin{center}
\includegraphics[scale=0.5, trim = 0mm 80mm 10mm 45mm, clip]{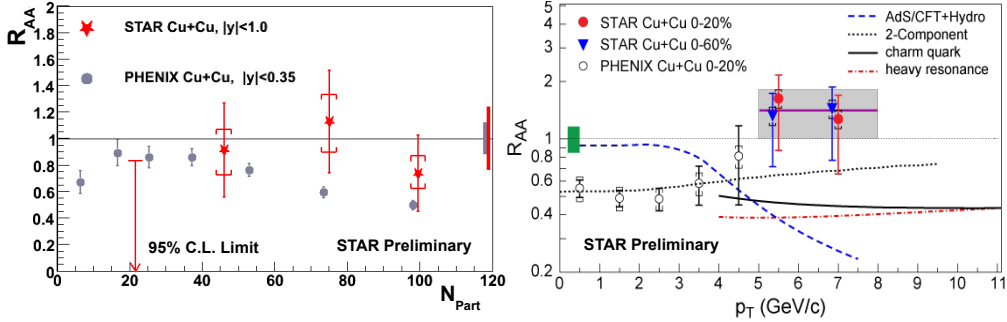} 
\caption{ \Jpsi\ \raa\ vs. centrality (\textit{left}) and $p_T$ (\textit{right}) for Cu+Cu at \sNN = 200GeV\cite{STARhighPtJpsi}. STAR
data points have statistical (bars) and systematic (caps)
uncertainties. The solid line and band on the left plot show the average and uncertainty of the two
0-20\% data points. The curves are model calculations described in
the text. The uncertainty band of 10\% for the dotted curve is not
shown. The boxes about unity on each plot show the $R_{AA}$
normalization uncertainty. }
\label{raa}
\end{center}
\end{figure}

\section{Data analysis and results}
The study of low-\pT\ \Jpsi\ production ($<$ 5 GeV/c) has been done using minimum bias data in Cu+Cu (year 2005) and Au+Au (year 2007) collisions at \sNN= 200 GeV while the high-\pT\ \Jpsi\ were studied using triggered data collected in Cu+Cu collisions. In the analysis reported here, the \Jpsi\ were reconstructed through the di-electron decay channel (branching ratio = 5.9\%). STAR is  well suited for such studies, as the large acceptance Time Projection Chamber (TPC) provides electron identification at moderate \pT\ (via dE/dx measurements), and at high-\pT\ the Barrel Electromagnetic Calorimeter (BEMC) is a very effective tool to identify electrons. The BEMC can be also used for fast online triggering to enrich the electron sample. 
In the case of minimum-bias Cu+Cu collisions (0-60\% of total cross-section, 27M events) only dE/dx from the STAR TPC was used to identify electrons and reject hadrons. For minimum-bias Au+Au collisions (0-80\% of total cross-section, 64M events), the BEMC provided an additional cut on the ratio of track momentum to energy. To improve the signal-to-background ratio, a cut of $p_T >$ 1.1 GeV/c in Cu+Cu ($p_T >$ 1.2 GeV/c in Au+Au) was applied for each electron selected in the studies.
For the high-\pT\ \Jpsi\ analysis, the BEMC triggered data for Cu+Cu collisions (year 2005) were used with transverse energy threshold  $E_T>$3.75 GeV (integrated luminosity  $\sim 860$ $\mu b^{-1}$). The higher-\pT\ electron is identified using combined information from the BEMC (tower energy), the Shower Max Detector embedded in BEMC (shower shape) and dE/dx measured in TPC, while only TPC is used to identify the second electron. To ensure a clean \Jpsi\ identification, \pT\ cuts of 3.5 GeV/c for the higher-\pT\ electron and 1.5 GeV/c for the lower-\pT\ electron were applied. 

Figure \ref{JpsiSignal} shows  the $e^{+}e^{-}$ invariant mass distributions in Au+Au at \sNN = 200 GeV. On the left plot, the signal and background distribution before background subtraction are presented for Au+Au, while on the right plot the \Jpsi\ invariant mass after background subtraction is shown. The background in the case of Au+Au, as well as for Cu+Cu collisions, was estimated by the event mixing technique and normalized to the total number of like -sign pairs. The shape of the \Jpsi\ mass peak is well described by a Crystal Ball function\cite{CrystalBallF}. Despite rather high background level (Signal-to-Background ratio = 1/13 for Cu+Cu and 1/28 for Au+Au) we have observed a prominent \Jpsi\ signal with significance of 4.5$\sigma$ in Cu+Cu and 6.5$\sigma$ in Au+Au collisions. The \Jpsi\ signal in Au+Au is promising and the studies are underway.

The \Jpsi\ \pT\ spectrum presented on Fig. \ref{pTspectrum} shows STAR's ability to measure the production of low and high \pT\ \Jpsi. The recent STAR upgrades, namely new Time-of-Flight detector and fast TPC DAQ system, will greatly improve the low-\pT\ \Jpsi\ measurements. The nuclear modification  factor (\raa) for \Jpsi\ as a function of centrality (represented by number of participants) (left panel) and \pT\ (right panel) for Cu+Cu at \sNN = 200 GeV are presented on Fig. \ref{raa} and compared to PHENIX results\cite{PhenixJpsiCuCu}. 
The \raa\ vs $N_{part}$ was obtained using minimum-bias Cu+Cu collisions only. The results are consistent with previously published data but, due to relatively modest statistics, the results have limited discrimination power. The \raa\ vs \pT\ for high-\pT\ \Jpsi\ is presented for top $20\%$  and $0-60\%$ Cu+Cu collisions\cite{STARhighPtJpsi}. The \raa\ seems to increase with \pT\ and it is consistent with unity for \pT\ $>$ 5 GeV/c: the average of the two STAR $0-20\%$ data points gives \raa\ = 1.4 $\pm$ 0.4 (stat.) $\pm$ 0.2 (syst.). The dashed curve shows the predictions of an AdS/CFT-based calculation embedded in a hydrodynamic model with hot wind dissociation incorporated - the data contradict the predicted \pT\ dependence. The dotted line represents calculations of the so called two component model\cite{ZaoRapp08} which includes \Jpsi\ dissociation, statistical   $c\overline{c}$ coalescence, \Jpsi\ formation time effects and B-meson feeddown. The model describes the overall trend of the data fairly well. The same calculations but without Bottom feeddown and finite formation time, predict decrease of \raa\ with \pT\ which may suggest the important role of these effects. On the other hand the recent STAR results show the B-meson feeddown contribution to \Jpsi\ production at high \pT\ is relatively small, on the level of 13\%\cite{STARhighPtJpsi}. The solid and dash-dot lines provide a comparison to open charm \raa. The solid line is based on the WHDG model\cite{WHDG_CharmQuark} for charm energy loss for $0-20\%$ Cu+Cu. The calculations include the elastic as well as radiative parton energy loss and the predictions were obtained with medium gluon density  $dN_g/dy\,=\,254$. The dash-dotted line are results of the GLV model\cite{GLV_HeavyResonance} for D-meson energy loss. In this approach the heavy meson suppression is described by collisional dissociation in quark-gluon plasma with $dN_g/dy\,=\,275$. Both calculations correctly describe heavy flavor suppression in Au+Au, thus one would expect a strong suppression for \Jpsi\ as well for \pT\ $>$ 5 GeV/c when they traverse through the medium carrying color charges, which is in contrast to the data. These results may suggest the high-\pT\ \Jpsi\ production is dominated by the color singlet channel but other processes, like Bottom feeddown or finite formation time, can compensate for the predicted suppression. It is worth to mention that the finite formation time, required to build the \Jpsi\ wave function, is expected to reduce the dissociation cross section for $\overline{c}c$ compared to fully formed charmonium. In case of longer formation time, some portion of \Jpsi\ is produced outside the medium and therefore it should be sensitive to gluons and charm quark energy loss in the quark-gluon plasma.

\section{Summary}

We reported recent studies of \Jpsi\ production in nucleus-nucleus collisions at STAR. The \Jpsi\ \raa\ seems to increase with \pT\ and we observed no suppression for \pT\ $>$ 5 GeV/c. The lack of \Jpsi\ suppression at high-\pT\ is in contrast to the predictions for substantial suppression of charm quarks and open charm\cite{WHDG_CharmQuark,GLV_HeavyResonance} at similar \pT, and this may suggest that \Jpsi\ in this \pT\ range is produced mostly via the color single channel.




\end{document}